\documentclass[aps,pra,reprint,amsmath,amssymb,superscriptaddress,super,twocolumn,longbibliography,nofootinbib,notitlepage]{revtex4-2}

\usepackage{mathtools}
\usepackage{siunitx}
\usepackage{hyperref}
\usepackage{caption}
\usepackage{subcaption}
\usepackage{titletoc}
\usepackage{tocloft}
\usepackage{multirow}
\usepackage{hhline}
\usepackage{tablefootnote}
\usepackage{makecell}

\usepackage[
    a4paper,
    left=0.7in,
    right=0.7in,
    top=0.7in,
    bottom=0.7in
]{geometry}
\begin{document}
\setcitestyle{super}
\title{\Large\bf Ultralow-power, high-speed programmable Si photonic circuits\\ with InGaAsP membrane}
\author{Tomohiro~Akazawa}
\thanks{akazawa@mosfet.t.u-tokyo.ac.jp}
\affiliation{Department of Electrical Engineering and Information Systems, The University of Tokyo, Bunkyo-ku, Tokyo 113-8656, Japan}

\author{Rui~Tang}
\affiliation{Department of Electrical Engineering and Information Systems, The University of Tokyo, Bunkyo-ku, Tokyo 113-8656, Japan}

\author{Hanzhi~Tang}
\affiliation{Department of Electrical Engineering and Information Systems, The University of Tokyo, Bunkyo-ku, Tokyo 113-8656, Japan}

\author{Makoto~Okano}
\affiliation{National Institute of Advanced Industrial Science and Technology (AIST), Tsukuba, Ibaraki 305-8569, Japan}

\author{Yangyang~Wan}
\affiliation{Department of Electrical Engineering and Information Systems, The University of Tokyo, Bunkyo-ku, Tokyo 113-8656, Japan}

\author{Nobuyuki~Matsuda}
\affiliation{Research Institute of Electrical Communication, Tohoku University, Aoba-ku, Sendai, Miyagi 980-8577, Japan}

\author{Kasidit~Toprasertpong}
\affiliation{Department of Electrical Engineering and Information Systems, The University of Tokyo, Bunkyo-ku, Tokyo 113-8656, Japan}

\author{Shinichi~Takagi}
\affiliation{Department of Electrical Engineering and Information Systems, The University of Tokyo, Bunkyo-ku, Tokyo 113-8656, Japan}

\author{Mitsuru~Takenaka}
\thanks{takenaka@mosfet.t.u-tokyo.ac.jp}
\affiliation{Department of Electrical Engineering and Information Systems, The University of Tokyo, Bunkyo-ku, Tokyo 113-8656, Japan}

\begin{abstract}
Programmable photonic circuits\cite{bogaerts2020programmable} have emerged as a promising platform for applications ranging from optical communications\cite{suzuki2019low,van2026real,wang2026all} to artificial-intelligence computing\cite{shen2017deep,bandyopadhyay2024single,pai2023experimentally,ahmed2025universal,ohno2022si,onodera2026arbitrary} and quantum information processing\cite{qiang2018large,maring2024versatile}, but their scaling is fundamentally constrained by their essential building block, the optical phase shifter. Existing phase-shifter technologies face inherent trade-offs among power consumption, operating speed, modulation efficiency, optical loss, and thermal crosstalk, making it challenging to realize high-performance, large-scale programmable photonic circuits. 
Here, we present a programmable photonic circuit based on InGaAsP/Si hybrid metal–oxide–semiconductor (MOS) phase shifters that combines ultralow power consumption, high-speed operation, high modulation efficiency, low optical loss and negligible thermal crosstalk. The phase shifters combine the low leakage current of a MOS capacitor with the strong carrier-induced refractive-index modulation of an InGaAsP membrane, achieving a static power consumption below 30 fW/$\pi$, a switching time of 555 ps, a phase-modulation efficiency ($V_\pi L$) of 0.13 Vcm and a carrier-induced excess insertion loss of only 0.20 dB/$\pi$. We integrate these phase shifters into a programmable Mach–Zehnder interferometer mesh and demonstrate optical switching and programmable unitary transformations, while maintaining femtowatt-level static power consumption across integrated phase shifters. We further demonstrate circuit-level operation with negligible thermal crosstalk, addressing a major obstacle to densely integrated programmable photonic circuits. These results establish a foundation for scalable, high-performance programmable photonic systems for next-generation signal processing and computation.
\end{abstract}

\maketitle

\section*{Introduction}
The ability to dynamically reconfigure the routing and processing of light has transformed photonic integrated circuits from fixed-function devices into versatile programmable systems\cite{bogaerts2020programmable}. Such programmable photonic circuits offer a common hardware platform for diverse applications, including signal processing\cite{van2026real,wang2026all}, optical switching\cite{suzuki2019low,yoo2021prospects,chu2023large,hu2025silicon,qiao201732}, machine-learning acceleration\cite{shen2017deep,bandyopadhyay2024single,pai2023experimentally,ahmed2025universal,ohno2022si,onodera2026arbitrary}, quantum information processing\cite{qiang2018large,maring2024versatile}, and sensing\cite{zhang2022large}. Advances in foundry-based silicon (Si) photonics have enabled a rapid increase in the integration density of photonic components\cite{siew2021review}. However, scaling programmable photonic circuits remains fundamentally constrained by the performance of an optical phase shifter, which is the essential building block of programmable photonic circuits. Existing phase-shifter technologies face inherent trade-offs among power consumption, operating speed, modulation efficiency, insertion loss and thermal crosstalk, arising from the materials and physical mechanisms used to induce optical phase shifts.

Thermo-optic (TO) phase shifters are among the most widely used phase shifters in programmable photonic circuits\cite{suzuki2019low,bandyopadhyay2024single,suzuki2017broadband}, because of their simple device structure and ease of integration. However, they typically require substantial static power ($>$ 1 mW) and exhibit relatively slow switching speeds ($>$ 10 $\mu$s) owing to the thermal nature of their operation\cite{suzuki2019low}. Heat diffusion also induces severe thermal crosstalk among neighboring phase shifters, complicating the independent control of densely integrated devices. These limitations pose major challenges to deploying TO phase shifters in large-scale programmable photonic circuits.
\begin{figure*}[t!]
    \centering
    \includegraphics[width=1.0\linewidth]{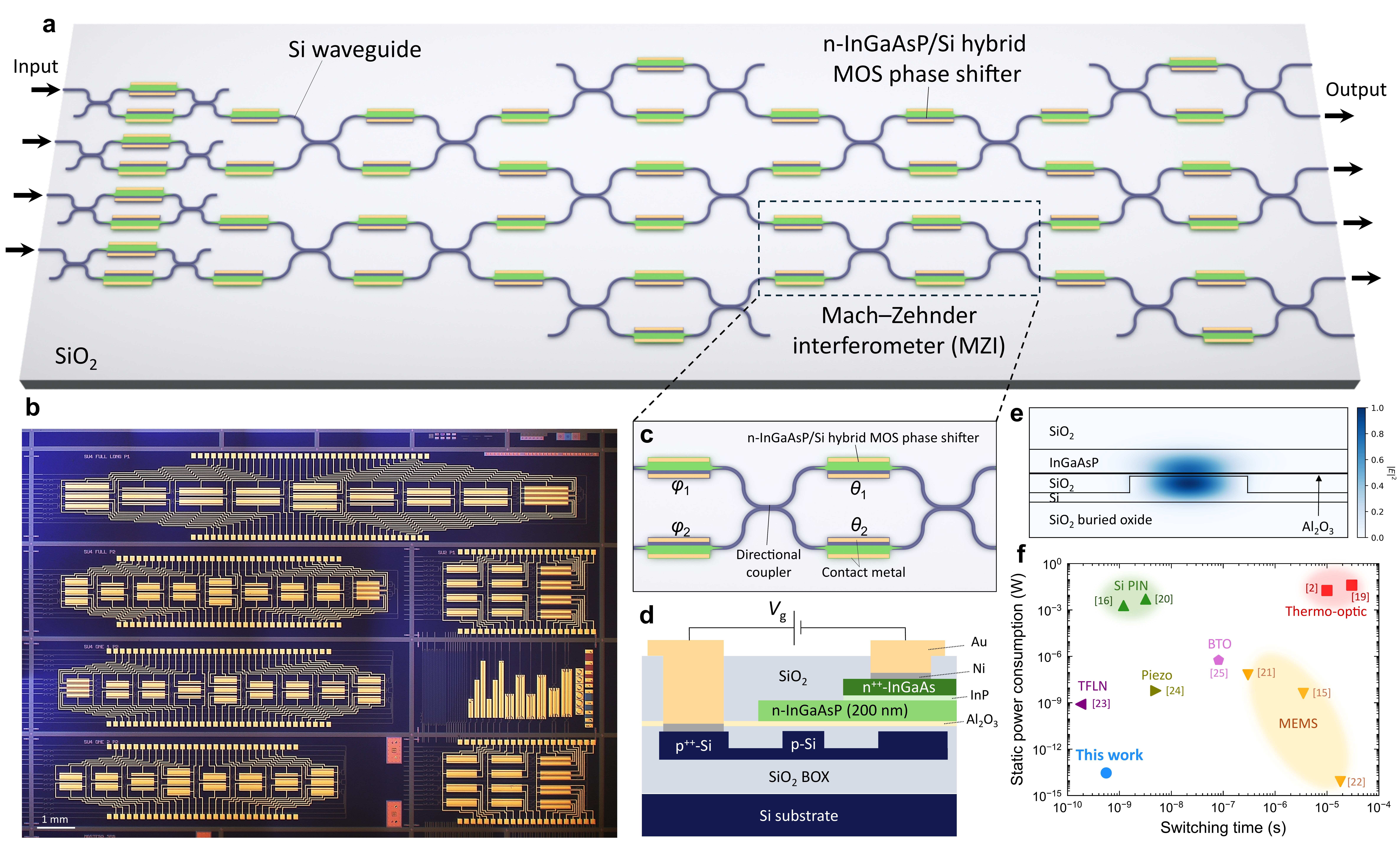}
    \caption{\textbf{Programmable photonic circuits with InGaAsP/Si hybrid MOS phase shifters.} (a) Schematic of a programmable photonic circuit incorporating InGaAsP/Si hybrid MOS phase shifters. (b) Optical microscopy image of a fabricated programmable photonic chip. (c) Building block of a programmable photonic circuit, consisting of a 2 $\times$ 2 Mach–Zehnder interferometer (MZI) with InGaAsP/Si hybrid MOS phase shifters. (d) Cross-sectional schematic of an InGaAsP/Si hybrid MOS phase shifter. (e) Fundamental transverse-electric (TE) mode profile of the InGaAsP/Si hybrid MOS phase shifter at a wavelength of 1550 nm. (f) Comparison of programmable photonic circuit platforms based on different phase-shifter technologies\cite{suzuki2019low,suzuki2017broadband,qiao201732,lu201616,seok2019wafer,kim2023programmable, hu2025silicon,zheng2024photonic,dong2022high,catala2026high} in terms of static power consumption and switching time.}
    \label{fig:figure1}
\end{figure*}
Alternatively, microelectromechanical systems (MEMS) phase shifters can achieve very low static power consumption\cite{hu2025silicon, seok2019wafer, kim2023programmable}; however, their mechanical actuation typically limits the switching speed ($>$ 1 $\mu$s) and require relatively high driving voltage ($>$ 10 V), posing challenges for high-speed operation and integration with electronic control circuits\cite{hu2025silicon}.
By contrast, Si optical phase shifters based on the free-carrier plasma-dispersion effect \cite{qiao201732, soref1987electrooptical, lu201616} offer nanosecond-scale switching speeds, but typically incur an excess insertion loss exceeding 1 dB\cite{qiao201732, lu201616} for a $\pi$ phase shift, owing to the large free-carrier absorption inherent to Si. This loss accumulates as light propagates through cascaded phase shifters, imposing a practical limitation on the scalability of programmable photonic circuits. Strain-based phase shifters using piezo-optomechanical actuators\cite{dong2022high} also exhibit nanosecond-scale switching speeds; however, their low modulation efficiency requires longer phase shifters, resulting in increased optical loss.
\begin{figure*}[t!]
    \centering
    \includegraphics[width=1.0\linewidth]{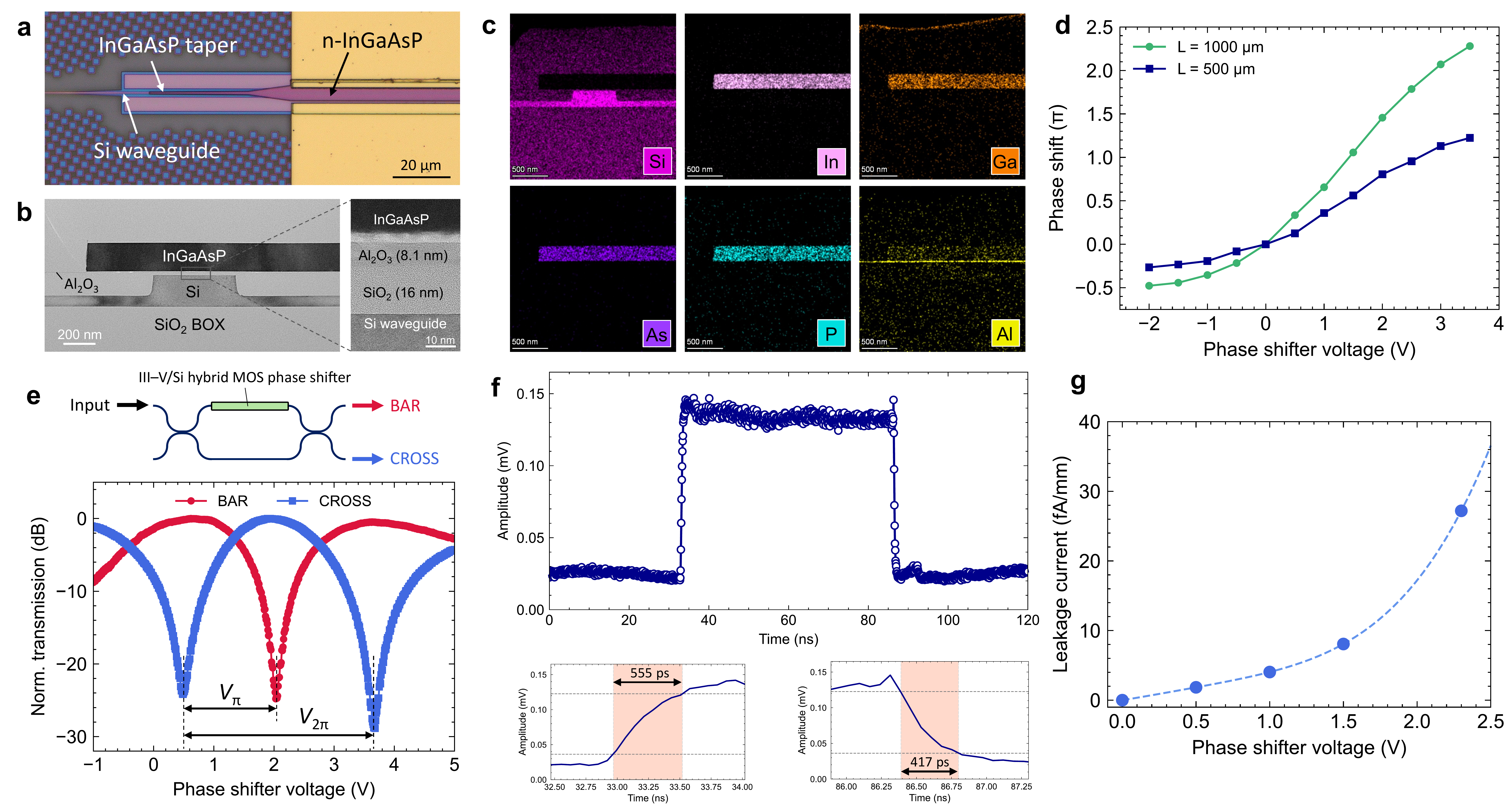}
    \caption{\textbf{Characterization of an n-InGaAsP/Si hybrid MOS optical phase shifter.} 
    (a) Optical microscopy image of an InGaAsP/Si hybrid MOS phase shifter. A two-step taper provides efficient mode transition between the Si rib waveguide and the InGaAsP/Si hybrid MOS phase shifter. (b) Cross-sectional TEM images and (c) EDS elemental maps of the InGaAsP/Si hybrid MOS phase shifter. (d) Phase shift as a function of applied gate voltage for phase-shifter lengths of 500 $\mu$m and 1000 $\mu$m. (e) Normalized transmission of a 2 $\times$ 2 MZI incorporating a 1.1-mm-long InGaAsP/Si hybrid MOS phase shifter as a function of gate voltage, demonstrating a phase-tuning range exceeding 2$\pi$. (f) Dynamic electro-optic response of a 2 $\times$ 2 MZI incorporating an InGaAsP/Si hybrid MOS phase shifter. The measured 10\% to 90\% rise and fall times are 555 ps and 417 ps, respectively. (g) Leakage current as a function of gate voltage. The static power for a $\pi$ phase shift is 28.9 fW for a 600-$\mu$m-long phase shifter.}
    \label{fig:figure2}
\end{figure*}
Thin-film lithium niobate (TFLN) phase shifters have been intensively studied in recent years\cite{zheng2024photonic, hu2025integrated,zhang2025thin} because of their ultrafast electro-optic response. However, they typically exhibit limited phase-modulation efficiency ($V_\pi L$ $>$ 1 Vcm)\cite{wang2018integrated}, reflecting the moderate Pockels coefficient of lithium niobate\cite{zhu2021integrated}. Moreover, TFLN phase shifters often rely on integrated heaters to stabilize the operating point\cite{holzgrafe2024relaxation}, introducing additional static power consumption and thermal crosstalk. Barium titanate (BTO), which exhibits a substantially stronger Pockels effect, has also attracted considerable attention for low-power, efficient, and high-speed phase shifting\cite{eltes2020integrated, kohli2023plasmonic}. A programmable photonic gate array based on BTO phase shifters has recently been demonstrated\cite{catala2026high}; however, the reported device consumed a static power exceeding 500 nW/$\pi$ and exhibited a phase-shifter insertion loss of 1.48 dB, indicating challenges associated with integrating BTO into Si photonic circuits. Furthermore, a switching time exceeding 80 ns was reported, which may be associated with the strongly dispersive dielectric response of BTO\cite{chelladurai2025barium}.

Further scaling of programmable photonic circuits therefore requires phase shifters that simultaneously achieve ultralow power consumption, high-speed operation, high modulation efficiency, low optical loss and thermal-crosstalk-free operation.

Here, we present an ultralow-power and high-speed programmable photonic circuit based on InGaAsP/Si hybrid metal-oxide-semiconductor (MOS) phase shifters, as illustrated in Fig. \ref{fig:figure1}a. We develop a fabrication process for integrating programmable Mach–Zehnder interferometer (MZI) meshes with n-InGaAsP/Si hybrid MOS phase shifters on a Si photonic platform (Fig. \ref{fig:figure1}b,c). The cross-sectional structure and fundamental transverse-electric (TE) mode profile of the hybrid MOS phase shifter are shown in Fig. \ref{fig:figure1}d,e, respectively. A 200-nm-thick n-InGaAsP membrane with a bandgap energy of 0.95 eV (corresponding to a bandgap wavelength of $\lambda_g$ =1300 nm) is used to suppress optical absorption at the operating wavelength of 1550 nm (Supplementary Section I). Under forward gate bias, electrons accumulate at the InGaAsP/Al$_2$O$_3$ interface and efficiently induce a negative refractive index change through the strong plasma dispersion and band-filling effects enabled by the small electron effective mass of n-InGaAsP (Supplementary Section II)\cite{han2017efficient,li2018ultra,hiraki2017heterogeneously, takenaka2019iii,bennett1990carrier,weber1994optimization,akazawa2025ultralow}. Under reverse gate bias, electron depletion and the Franz–Keldysh effect contribute to a positive refractive index change\cite{li2020optical}. This bidirectional refractive index tuning mechanism enables a wide phase-tuning range. The hybrid MOS architecture in this work achieves a high phase-modulation efficiency of 0.13 Vcm with a carrier-induced excess loss of only 0.20 dB/$\pi$. The ultralow leakage current of the SiO$_2$-embedded MOS structure reduces the static power consumption to 28.9 fW/$\pi$ (Supplementary Section III), while a switching time of 555 ps is experimentally demonstrated. A comparison with previously reported programmable photonic circuit platforms\cite{suzuki2019low, hu2025silicon, qiao201732,suzuki2017broadband, kim2023programmable, seok2019wafer, lu201616, dong2022high, zheng2024photonic, catala2026high} shows that our platform combines femtowatt-level static power consumption with sub-nanosecond switching (Fig. \ref{fig:figure1}f), together with high modulation efficiency, low optical loss and thermal-crosstalk-free operation (see Table \ref{tab:table1} for a comprehensive comparison). We construct a programmable MZI mesh using the InGaAsP/Si hybrid MOS phase shifters and demonstrate high-fidelity optical switching and programmable unitary transformations, as well as circuit-level operation without thermal crosstalk. We further assess the scalability of the ultralow-power operation by measuring the leakage current of every phase shifter in the programmable photonic circuit, finding femtowatt-level static power consumption for the vast majority of phase shifters. These results provide a promising pathway towards scaling programmable photonic circuits beyond the conventional trade-offs of phase-shifter technologies.

\section*{Characterization of InGaAsP/Si hybrid MOS phase shifter}

InGaAsP/Si hybrid MOS phase shifters were fabricated by direct wafer bonding\cite{han2016study} of n-InGaAsP/n-InP/n$^{++}$-InGaAs epitaxial layers grown on an InP substrate onto SiO$_2$-embedded p-type Si rib waveguides using an Al$_2$O$_3$ bonding layer (Methods and Supplementary Section III). Figure \ref{fig:figure2}a shows an optical microscopy image of an InGaAsP/Si hybrid MOS phase shifter, in which a two-step taper provides an efficient mode transition between the Si rib waveguide and the InGaAsP/Si hybrid waveguide (Supplementary Section IV). Cross-sectional transmission electron microscopy (TEM) images (Fig. \ref{fig:figure2}b) reveal the hybrid MOS structure, with an equivalent oxide thickness (EOT) of 19.2 nm determined from the magnified image. Energy-dispersive X-ray spectroscopy (EDS) elemental maps (Fig. \ref{fig:figure2}c) further confirm the successful bonding of the InGaAsP phase-shifter layer onto the Si rib waveguide through the Al$_2$O$_3$ bonding layer.
\begin{figure*}[t!]
    \centering
    \includegraphics[width=1.0\linewidth]{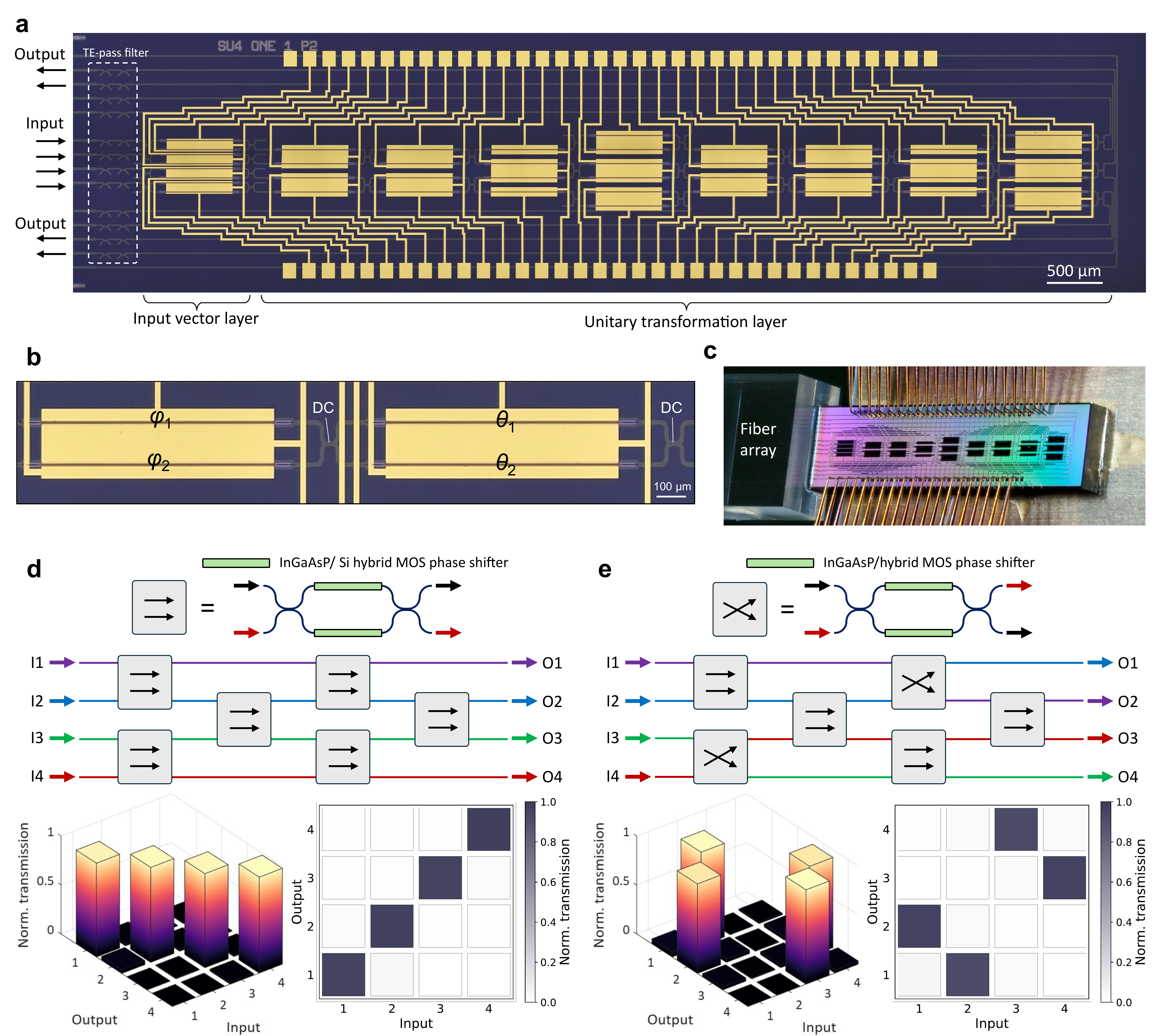}
    \caption{\textbf{Optical switching with a programmable photonic circuit based on InGaAsP/Si hybrid MOS phase shifters.}  (a) Optical microscopy image of the fabricated 4 $\times$ 4 programmable photonic circuit consisting of an input vector layer and a unitary transformation layer implemented with InGaAsP/Si hybrid MOS optical phase shifters. (b) Optical microscopy image of a 2 $\times$ 2 Mach–Zehnder interferometer (MZI), the fundamental building block of the programmable photonic circuit. The internal ($\theta_1$ and $\theta_2$) and external ($\varphi_1$ and $\varphi_2$) phase-shifter pairs are operated in a push–pull configuration. (c) Photograph of the packaged programmable photonic circuit used for the measurements. (d)(e) Optical routing configurations and corresponding measured normalized transfer matrices for two representative circuit configurations. The MZIs are programmed into the bar or cross states to realize the desired input–output routing.}
    \label{fig:figure3}
\end{figure*}
We first characterized the phase-modulation efficiency using asymmetric MZIs (AMZIs) incorporating 500-$\mu$m- and 1000-$\mu$m-long InGaAsP/Si hybrid MOS phase shifters. Figure \ref{fig:figure2}d shows the phase shift extracted from the transmission spectra of the AMZI as a function of gate voltage. A high phase-modulation efficiency, with a $V_\pi L$ of 0.13 Vcm, is achieved owing to the small effective electron mass of InGaAsP. The carrier-induced excess loss for a $\pi$ phase shifter is only 0.20 dB (Supplementary Section V). Including the two Si-to-hybrid-waveguide tapers, each with an insertion loss of 0.18 dB (Supplementary Section IV), the total insertion loss of the phase-shifter section is 0.56 dB. Further optimization of the fabrication process reduces the taper loss to below 0.01 dB per taper, indicating that the total insertion loss can potentially be reduced to below 0.22 dB (Supplementary Section IV). Figure \ref{fig:figure2}e shows the transmission spectra of the bar and cross ports of a 2 $\times$ 2 MZI incorporating a 1.1-mm-long InGaAsP/Si hybrid MOS phase shifters at a wavelength of 1550 nm as the gate voltage is swept from $-$1 to 5 V, demonstrating a phase-tuning range exceeding 2$\pi$. Figure \ref{fig:figure2}f shows the dynamic optical response of the MZI driven by 9.375-MHz square electrical pulses with a peak-to-peak voltage ($V_\mathrm{pp}$) of 2.0 V and a DC bias of 3.0 V. The measured rise and fall times are 555 ps and 417 ps, respectively, demonstrating sub-nanosecond switching. The switching speed is currently limited by the RC time constant of the phase shifter and could be further improved by reducing parasitic capacitance \cite{tang2022modulation} (Supplementary Section VI).

Figure \ref{fig:figure2}g shows the current–voltage ($I$--$V$) characteristics of the phase shifter. Because of the extremely small leakage current, for which the contribution of capacitive current cannot be neglected, we employed a current-sampling method to accurately determine the steady-state leakage current (Supplementary Section VII). For the 600-$\mu$m-long phase shifter used in the programmable photonic circuit, the leakage current at a $\pi$ phase shift was measured to be 13.3 fA. This corresponds to a static power consumption of only 28.9 fW/$\pi$, approximately ten orders of magnitude lower than that of conventional TO phase shifters. This ultralow static power consumption originates from the extremely low leakage current of the MOS structure, aided by the dense embedded SiO$_2$ cladding surrounding the Si waveguide (Supplementary Section III). We further measured the capacitance of the InGaAsP/Si hybrid MOS phase shifter to evaluate the dynamic switching energy (Supplementary Section VI). The capacitance was 6.48 pF at a $V_\pi$ = 2.2 V for a 600-$\mu$m-long phase shifter, corresponding to a switching energy of 15.3 pJ/$\pi$. The switching energy could be further reduced by decreasing the parasitic capacitance and gate-oxide thickness of the MOS phase shifter (see Discussion for details).

\section*{Programmable photonic circuit with InGaAsP/Si hybrid MOS phase shifters}

We characterized a 4 $\times$ 4 programmable photonic circuit comprising a four-mode input vector layer and a unitary-transformation layer\cite{harris2018linear, miller2013self} (Fig. \ref{fig:figure3}a). The input vector layer consists of four MZIs that independently control the amplitude of the input light in each mode. The unitary-transformation layer is implemented using a Clements mesh\cite{clements2016optimal} of cascaded 2 $\times$ 2 MZIs, each comprising directional couplers (DCs), internal phase shifters ($\theta_1$ and $\theta_2$) and external phase shifters ($\varphi_1$ and $\varphi_2$), as shown in Fig. \ref{fig:figure3}b. The internal and external phase-shifter pairs are operated in a push–pull configuration, providing a wide phase-tuning range at reduced driving voltages. Figure \ref{fig:figure3}c shows the photograph of the packaged programmable photonic circuit under measurement. An optical fiber array is attached to the photonic circuit, and the phase-shifter voltages are independently controlled using a multichannel source-measure unit (SMU) (See Methods for details).

 We first programmed the photonic circuit to implement different switching configurations at a wavelength of 1550 nm. By setting the phase difference between the two arms of each MZI to either 0 (cross state) or $\pi$ (bar state), the input optical signals can be routed to different output ports, enabling reconfigurable input–output mapping. Figures \ref{fig:figure3}d,e show representative measured normalized transfer matrices for two switching configurations, implemented by driving only the internal phase shifters of each MZI. The measured transfer matrices closely reproduce the programmed switching patterns, with high transmission at the targeted output ports and strong suppression at the undesired ports. A channel crosstalk below $-$12 dB is achieved, demonstrating reconfigurable optical signal routing using the InGaAsP/Si hybrid MOS programmable photonic circuit.
 \begin{figure*}[t!]
    \centering    \includegraphics[width=1.0\linewidth]{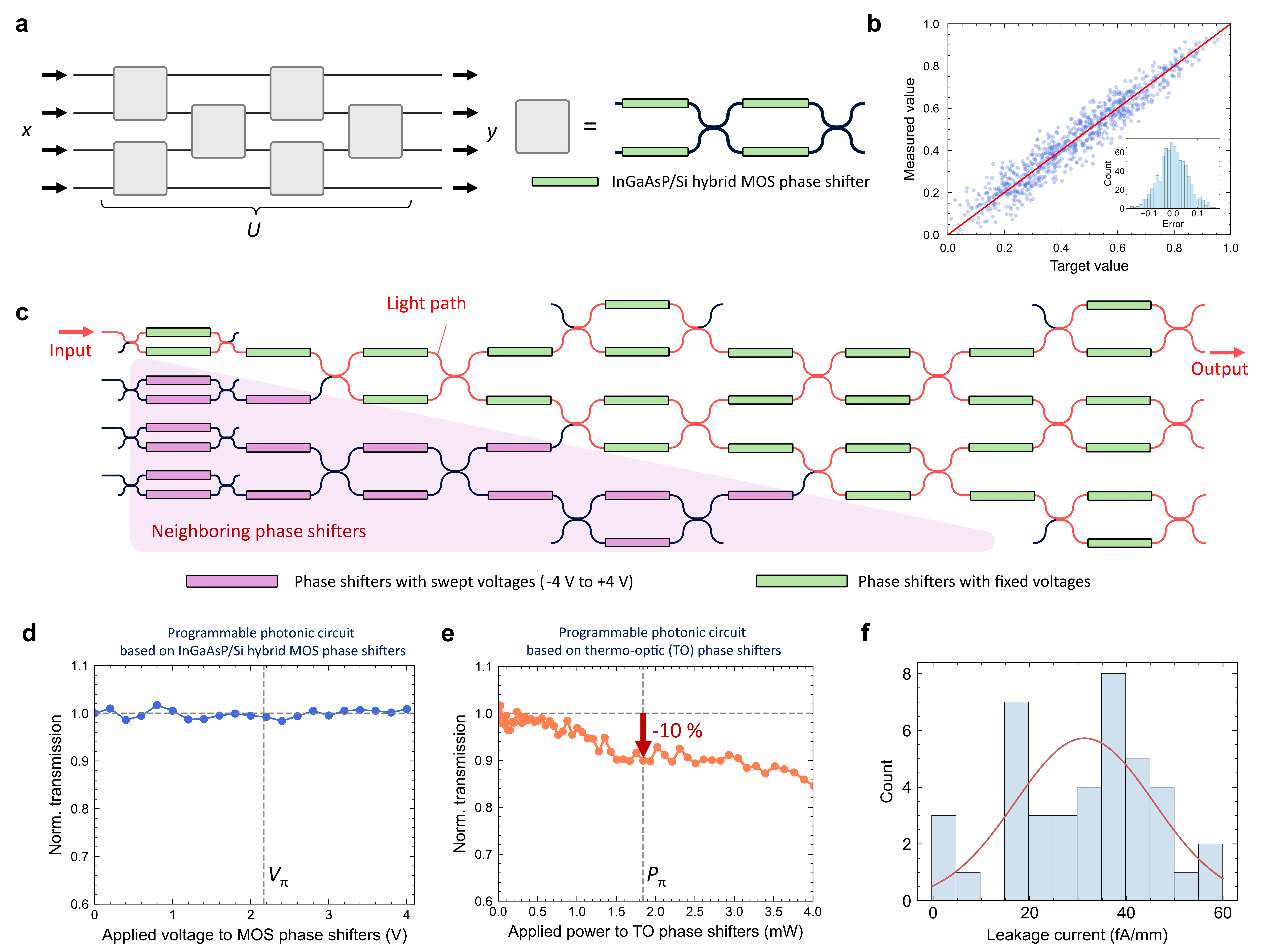}
    \caption{\textbf{Linear transformations and thermal-crosstalk-free operation of the programmable photonic circuit.}  
      (a) Schematic of the 4 $\times$ 4 programmable photonic linear transformer based on a Clements mesh\cite{clements2016optimal}. (b) Comparison between the target and measured normalized transmission values for 50 randomly generated unitary matrices. The inset shows the distribution of errors between the target and measured values. (c) Experimental scheme for evaluating thermal crosstalk in the programmable photonic circuit. The voltages applied to neighboring phase shifters outside the optical path are varied while monitoring the transmission through the programmed optical path. (d) Normalized transmission through the programmable photonic circuit based on InGaAsP/Si hybrid MOS phase shifters as a function of the voltage applied to the neighboring phase shifters. The transmission remains nearly unchanged over the entire voltage range, demonstrating negligible thermal crosstalk. (e) Normalized transmission through a programmable photonic circuit based on TO phase shifters as a function of the electrical power applied to the neighboring phase shifters, showing a substantial transmission change induced by thermal crosstalk. (f) Distribution of leakage currents measured across the InGaAsP/Si hybrid MOS phase shifters in the programmable photonic circuit.}
    \label{fig:figure4}
\end{figure*}
 To demonstrate the capability of our programmable photonic circuit as a linear photonic processor, we implemented 50 randomly generated unitary matrices using the InGaAsP/Si hybrid MOS phase shifters. To compensate for phase errors arising from fabrication imperfections, we employed the covariance matrix adaptation evolution strategy (CMA-ES) to globally optimize the phase-shifter voltages for each target unitary matrix\cite{zhang2021efficient}(Supplementary Section VIII). During the optimization, the voltage range was restricted to $-$4 to 4 V, and both the internal and external phase shifters of each MZI in the Clements mesh were controlled to realize the target transformations, as illustrated in Fig. \ref{fig:figure4}a. Figure \ref{fig:figure4}b compares the target and measured normalized transmission values for 50 randomly generated unitary matrices. The measured values closely follow the target values, with a coefficient of determination ($R^2$) of 0.93 and a standard deviation ($\sigma$) of 0.052 for the error distribution (inset of Fig. \ref{fig:figure4}b). These results demonstrate that diverse unitary transformations can be implemented with high fidelity using our programmable photonic circuit based on InGaAsP/Si hybrid MOS phase shifters.
 
 We next investigated thermal crosstalk in the programmable photonic circuit based on InGaAsP/Si hybrid MOS phase shifters. Thermal crosstalk arises when heat generated by an actively driven phase shifter unintentionally perturbs the optical phase in neighboring devices. This effect poses a major challenge to scaling programmable photonic circuits based on TO phase shifters, which typically consume more than 1 mW of static power and generate substantial heat during operation. By contrast, the femtowatt-level static power consumption of our InGaAsP/Si hybrid MOS phase shifters is expected to essentially eliminate thermal crosstalk. To experimentally verify this, continuous-wave (CW) light at a wavelength of 1550 nm was injected into the top input of the programmable photonic circuit along the optical path indicated in red in Fig. \ref{fig:figure4}c. The same gate voltage, swept from 0 to 4 V, was simultaneously applied to all neighboring phase shifters outside the optical path, while the phase shifters along the optical path were maintained at fixed voltages. The experiment was performed without a thermoelectric cooler (TEC) and therefore without active temperature stabilization. As shown in Fig. \ref{fig:figure4}d, the optical transmission remains nearly unchanged over the entire voltage range, demonstrating negligible thermal crosstalk at the circuit level. For comparison, we performed the same experiment using a programmable photonic circuit based on TO phase shifters (Supplementary Section IX). In this case, the transmission changes markedly with the electrical power applied to the neighboring TO phase shifters, exceeding 10\% at the power required for a $\pi$ phase shift ($P_\pi$), as shown in Fig. \ref{fig:figure4}e. This direct comparison demonstrates that the ultralow static power consumption of the InGaAsP/Si hybrid MOS phase shifters effectively eliminates the thermal crosstalk inherent to TO-based programmable photonic circuits.
 
 Finally, to assess the scalability of the ultralow-power operation at the circuit level, we measured the leakage current of every phase shifter in a programmable photonic circuit at a bias voltage of 2.3 V, approximately corresponding to $V_\pi$ for a 600-$\mu$m-long phase shifter. As shown in Fig. \ref{fig:figure4}f, the leakage currents remain extremely low across the circuit, with the vast majority of phase shifters exhibiting femtowatt-level static power consumption and only a few exceeding 1 pW. The median static power consumption normalized by the phase-shifter length is only 36.4 fW mm$^{-1}$, corresponding to 21.8 fW for a 600-$\mu$m-long phase shifter. These results demonstrate that femtowatt-level static power consumption is consistently maintained across the programmable photonic circuit, supporting the scalability of the InGaAsP/Si hybrid MOS platform towards larger photonic circuits.

\begin{table*}[t!]
\centering
\caption{\textbf{Comparison of state-of-the-art programmable photonic circuit platforms with different phase-shifter technologies based on circuit-level demonstrations.}}
\renewcommand{\arraystretch}{2}
\begin{tabular*}{\textwidth}{@{\extracolsep{\fill}}cccccccc@{}}
\hline
Phase shifter &
\makecell{Phase-tuning\\mechanism} &
$V_\pi L$ &
\makecell[c]{Static\\power} &
\makecell[c]{Switching\\speed} &
\makecell{Switching\\energy} &
\makecell{Insertion\\loss} &
\makecell{Thermal\\crosstalk} \\
\hline\hline
\makecell{Thermo-\\optic\cite{suzuki2019low}} & \makecell{Thermo-optic\\effect} & N.A. & 18.1 mW/$\pi$& 33.3 $\mu$s&$^{(3)}$603 nJ/$\pi$ &N.A. & Large\\ \hline
MEMS\cite{kim2023programmable} &Optomechanical & 0.075 Vcm & 8.6 fW/$\pi$& 18.1 $\mu$s&24.5 pJ/$\pi$ &0.13 dB & $^{(4)}$None\\ \hline
Si PIN\cite{qiao201732} & \makecell{Carrier effect\\ in Si} & N.A. & 6.24 mW/$\pi$& 1.2 ns&$^{(3)}$7.5 pJ/$\pi$ &$^{(5)}$1.3 dB & Moderate\\ \hline
$^{(1)}$Piezo\cite{dong2022high} & \makecell{Piezo-\\optomechanical} & 50 Vcm & 6 nW/$\pi$& 5.0 ns&N.A. &3.5 dB & $^{(4)}$None\\ \hline
TFLN\cite{zheng2024photonic} &Pockels effect & 3.7 Vcm & $^{(2)}$800 pW/$\pi$& 180 ps&$^{(4)}$403 fJ/$\pi$ &N.A. & $^{(4)}$None\\ \hline
BTO\cite{catala2026high} & Pockels effect & 0.18 Vcm &560 nW/$\pi$& 81.5 ns&$^{(4)}$44.8 fJ/$\pi$ &1.48 dB & $^{(4)}$None\\ \hline
\makecell{InGaAsP/Si\\hybrid MOS\\(this work)
} & \makecell{Carrier effect\\in InGaAsP} & 0.13 Vcm & 28.9 fW/$\pi$& 555 ps&15.3 pJ/$\pi$ &$^{(5)}$0.56 dB & None\\ 
 \hline
\end{tabular*}
\vspace{2pt} \parbox{\textwidth}{
\raggedright \footnotesize 
$^{(1)}$ The operating wavelength is 737 nm. \\ 
$^{(2)}$ Includes static power required to stabilize the operating point.\\ 
$^{(3)}$ Calculated using $E_\pi = P_\pi t_\mathrm{switch}$.\\ 
$^{(4)}$ Experimental data are not shown.\\
$^{(5)}$ The insertion loss at a $\pi$ phase shift is shown.}
\label{tab:table1}
\end{table*}

\section*{Discussion and Outlook}
We have demonstrated an ultralow-power, high-speed programmable photonic circuit platform based on InGaAsP/Si hybrid MOS phase shifters. By combining the extremely low leakage current of the MOS structure with the strong carrier-induced refractive-index modulation of the InGaAsP membrane, our platform simultaneously achieves femtowatt-level static power consumption, sub-nanosecond switching, high phase-modulation efficiency and low optical loss. The phase shifters exhibit a static power consumption below 30 fW/$\pi$ and a switching time of 555 ps. The small effective electron mass of InGaAsP enables efficient carrier-induced phase modulation, resulting in a $V_\pi L$ of 0.13 Vcm with a carrier-induced excess loss of only 0.20 dB/$\pi$. Including the two Si-to-hybrid-waveguide tapers, the total insertion loss of the phase-shifter section is 0.56 dB.

Table \ref{tab:table1} compares key figures of merit of phase shifters demonstrated in state-of-the-art programmable photonic circuit platforms\cite{suzuki2019low, kim2023programmable, qiao201732, dong2022high, zheng2024photonic, catala2026high}. A key achievement of our platform is that it simultaneously achieves high phase-modulation efficiency, high-speed switching, ultralow static power consumption, low switching energy, low insertion loss and negligible thermal crosstalk, thereby overcoming the trade-offs among these metrics that have constrained the scaling of programmable photonic circuits. Conventional programmable photonic circuits based on TO phase shifters typically require more than 1 mW of static power per phase shifter, leading to substantial circuit-level power consumption and thermal crosstalk as the number of phase shifters increases. By contrast, the InGaAsP/Si hybrid MOS phase shifters consume only femtowatt-level static power. Based on the median static power consumption of 21.8 fW per phase shifter, the total phase-shifter static power of a circuit incorporating one million phase shifters would be only $\sim$22 nW, substantially relaxing the power and thermal constraints on large-scale integration. Moreover, unlike MEMS phase shifters, whose switching speed is fundamentally limited by mechanical actuation, our phase shifters achieve sub-nanosecond switching owing to their small RC time constant. This fast response is particularly advantageous for applications requiring rapid circuit reconfiguration, such as optical packet switching (OPS).

However, the present phase shifter design exhibits a switching energy of 15.3 pJ/$\pi$, which is relatively high compared with those of phase shifters based on the Pockels effect, such as TFLN and BTO. There are, however, clear engineering pathways to substantially reduce this energy. The switching energy for a $\pi$ phase shift ($E_\pi$) scales approximately as $E_\pi = C_\mathrm{ox}V_\pi^2/2$, where $C_\mathrm{ox}$ is the MOS capacitance and $V_\pi$ is the voltage required for a $\pi$ phase shift. Two approaches can therefore be used to reduce the switching energy. First, the parasitic capacitance can be reduced by removing Si slab regions that do not contribute to phase modulation (Supplementary Section VI). Second, reducing the EOT ($t_\mathrm{ox}$) of the MOS phase shifter by a factor of $\kappa$ increases $C_\mathrm{ox}$ by a factor of $\kappa$ while reducing $V_\pi$ by the same factor, resulting in an overall $\kappa$-fold reduction in $E_\pi$. Combining the removal of parasitic capacitance with a tenfold reduction in EOT is expected to reduce the switching energy by approximately 30-fold, from 15.3 pJ/$\pi$ to $\sim$500 fJ/$\pi$.

The InGaAsP/Si hybrid MOS phase shifter demonstrated in this work exhibits a low insertion loss of 0.56 dB, comprising a carrier-induced excess loss of 0.20 dB/$\pi$ and an insertion loss of 0.18 dB for each of the two Si-to-hybrid waveguide tapers. This total loss is substantially lower than those reported for Si PIN and BTO phase shifters in the circuit-level demonstrations ($>$ 1 dB; Table 1). The insertion loss can be further reduced by optimizing the taper fabrication. The present taper loss of 0.18 dB is primarily attributed to the relatively large taper-tip width resulting from fabrication imperfections. Reducing the taper-tip width to 150 nm is expected to decrease the insertion loss to below 0.01 dB per taper (Supplementary Section IV), potentially reducing the total insertion loss of the phase-shifter section to approximately 0.22 dB.

Based on the InGaAsP/Si hybrid MOS phase shifters, we have demonstrated key functionalities of programmable photonic circuits, including optical switching and unitary transformations, highlighting the potential of this platform for applications ranging from optical interconnects in data centers to deep-learning accelerators and quantum information processing. Importantly, the femtowatt-level static power consumption of the phase shifters enables circuit operation without thermal crosstalk, as experimentally demonstrated in this work. This capability addresses a major limitation of TO-based programmable photonic circuits, in which the substantial static power consumption of the phase shifters leads to thermal crosstalk and increasingly complex control as the circuit scales. Together, these results establish the InGaAsP/Si hybrid MOS platform as a promising foundation towards large-scale programmable photonic circuits combining versatile functionality with ultralow-power operation.

In conclusion, we have established a programmable photonic circuit platform based on InGaAsP/Si hybrid MOS phase shifters. By integrating n-type InGaAsP membranes with Si photonic circuits, our platform overcomes the conventional trade-offs among power consumption, operating speed, modulation efficiency, optical loss and thermal crosstalk that have constrained the scaling of programmable photonic circuits. Although further optimization of the device design and fabrication is required to reduce insertion loss and switching energy, the simultaneous achievement of femtowatt-level static power consumption, sub-nanosecond switching and thermal-crosstalk-free circuit operation provides a scalable route towards large-scale programmable photonic systems for both classical and quantum applications.

\section*{Methods}

\subsection*{Device fabrication}
Si waveguides were first fabricated on a 300-mm silicon-on-insulator (SOI) wafer and subsequently embedded in SiO$_2$ by thermal oxidation and chemical vapor deposition (CVD). The wafer surface was then planarized by chemical mechanical polishing (CMP). III--V epitaxial layers, including 200-nm-thick n-InGaAsP ($\lambda_g = $1300 nm), 20-nm-thick n-InP, and 100-nm-thick n$^{++}$-InGaAs layers, were bonded to a SOI chip containing the SiO$_2$-embedded Si waveguides using an 8-nm-thick Al$_{2}$O$_{3}$ layer. The n$^{++}$-InGaAs and n-InP regions were subsequently defined by electron-beam (EB) lithography and selective wet etching. The InGaAsP regions forming the phase shifters were patterned by dry etching, followed by deposition of a 610-nm-thick SiO$_2$ cladding layer for surface passivation. Finally, Ni/Au metal contacts were formed by electron-beam evaporation and lift-off.
\subsection*{Measurement}
For optical characterization of individual devices, continuous-wave (CW) light from a tunable laser (Santec, TSL-510) was coupled into the Si waveguide through a Si edge coupler using a single-mode optical fiber. The polarization of the input light was adjusted to excite the transverse-electric (TE) mode of the Si waveguide using an in-line polarization controller. The output light was coupled from the Si waveguide into an optical fiber through another edge coupler and measured using an optical power meter (Santec, MPM-210H). A precision source/measure unit (SMU; Keysight, B2900A) was used to apply the bias voltage. The leakage current and capacitance of the phase shifter were measured using a semiconductor device analyzer (Keysight, B1500A). For dynamic characterization, CW light was coupled into the device and square-wave electrical pulses were applied to the phase shifter using a pulse pattern generator (PPG; Anritsu, MU181020A). The modulated optical signals were recorded using an oscilloscope (Keysight, 86100D).
For circuit-level characterization, the fabricated chip was first diced by stealth dicing, and the chip facets were subsequently polished. A high-numerical-aperture (NA) fiber array was attached to the chip for optical input and output. CW light was routed to the desired input port using a MEMS optical switch. The phase-shifter voltages were independently controlled using a multichannel SMU (Nicslab, XDAC-120MUB-R4G8), and the optical powers at the output ports were simultaneously measured using a multichannel high-speed optical power meter (OptoTest Corp., OP760). The packaged chip was photographed using a mirrorless camera with a macro lens (OM System OM-5, M.ZUIKO DIGITAL ED 60mm F2.8 Macro).

\section*{Acknowledgements}
We thank E. Kato, T. Tanemura, and Y. Nakano for support with device fabrication. This work was supported by JST CREST (JPMJCR2004), and the Japan Society for the Promotion of Science (JP23H00172, JP24KJ0823). Part of this work was conducted at the Takeda Sentanchi super cleanroom, The University of Tokyo, supported by the Nanotechnology Platform Program of the Ministry of Education, Culture, Sports, Science and Technology (MEXT), Japan (JPMXP1224UT1028).

\section*{Author contributions}
T.A. and M.T. conceived the project. T.A. designed the devices and photonic circuits with assistance from R.T., H.T. and N.M. T.A. performed the III--V integration onto the Si photonic circuits. T.A. performed measurements of individual devices and programmable photonic circuits with assistance from R.T. and W.Y. T.A. analyzed the results and prepared the figures. T.A. and M.T. wrote the manuscript. K.T. and S.T. contributed to the overall discussion. M.T. contributed to manuscript revision and provided overall project supervision.

\section*{Competing interests}
The authors declare no competing interests.

\bibliography{references}

\end{document}